\documentclass[pra,superscriptaddress,amsmath,twocolumn]{revtex4}
\usepackage{graphicx}
\usepackage{epstopdf}
\usepackage{color}

\newcommand{\9}{\rangle}

\newcommand{\bra}[1]{\langle #1 |}
\newcommand{\ket}[1]{| #1 \rangle}
\newcommand{\braket}[2]{\langle #1 | #2 \rangle}

\newcommand{\half}{\mbox{$1\over2$}}
\newcommand{\phalf}{\mbox{$\varphi\over2$}}
\newcommand\tr{{\mbox{tr\,}}}

\newcommand{\etal}{{  et al.}}
\newcommand{\be}{\begin{equation}}
\newcommand{\ee}{\end{equation}}
\newcommand{\ba}{\begin{eqnarray}}
\newcommand{\ea}{\end{eqnarray}}

\newcommand\V{{\cal V}}
\newcommand\p{{\sf p}}
\newcommand\w{{\sf w}}

\begin{document}

\title{Quantum control in foundational experiments}


\author{Lucas C.  C\'{e}leri}
\author{Rafael M. Gomes}
\affiliation{Instituto de F\'{\i}sica, Universidade Federal de Goi\'{a}s, Goi\^{a}nia, GO, Brazil}
\author{Radu Ionicioiu}
\affiliation{Department of Theoretical Physics, National Institute of Physics and Nuclear Engineering,
Bucharest--M\u{a}gurele, Romania}
\author{        Thomas Jennewein }
\affiliation{Institute for Quantum Computing University of Waterloo, Waterloo ON, Canada}
 \affiliation{Department of Physics and Astronomy, University of Waterloo, Waterloo ON, Canada}
\author{       Robert B. Mann  }
 \affiliation{Department of Physics and Astronomy, University of Waterloo, Waterloo ON, Canada}
 \affiliation{Perimeter Institute for Theoretical Physics,  Waterloo ON, Canada}
\author{        Daniel R. Terno}
\affiliation{Department of Physics and Astronomy, Macquarie University, Sydney NSW, Australia}
             \email{daniel.terno@mq.edu.au}


\begin{abstract}
We describe a new class of experiments designed to probe the foundations of quantum mechanics.  Using quantum controlling devices, we show how to attain a freedom in temporal ordering of the control and detection of various phenomena. We consider wave-particle duality in the context of quantum-controlled and the entanglement-assisted delayed-choice experiments. Then we discuss a quantum-controlled CHSH experiment and measurement of photon's transversal position and momentum in a single set-up.

\end{abstract}

\maketitle

\section{Introduction}
\label{intro}

 The   Bohr-Einstein discussions  on the nature of quantum theory \cite{wz,e-c} were responsible for the appearance of the first modern \textit{gedanken} experiments. These thought experiments became the  weapons of choice in the struggle of our classical intuition with quantum mechanics. In the last decades they developed into common lab procedures. Former paradoxes of quantum foundations  are now  resources of quantum information  science \cite{nc}.
This  new technological ability allows refining of now classic  experiments \cite{compendum,aspect82,gra86,longhi}, as well as probing other aspects of quantum foundations.

  Wave-particle duality, superposition and entanglement are just   some of the quantum concepts that run afoul of our classical expectations. Hidden-variable (HV) theories  are proposed to remove or explain these non-classical features.  Moreover, an additional set of rules (measurement description) draws quantum possibilities into an irreversible classical record \cite{compendum,peres}.  This happens despite measuring devices being built from quantum constituents.

In the von Neumann's discussion of measurement \cite{vN} a quantum system is used to observe the preceding one, until the chain of systems is cut by a classical observer (or a device).  Keeping one link in this chain  makes quantum   controlling devices to perform the switching between different classical set-ups.
The first example of  a quantum control is a radioactive atom in the Schr\"{o}dinger's cat \emph{gedankenexperiment} \cite{cat}. By correlating decayed and undecayed states of an atom with dead and alive states of a cat, it demonstrated non-classical properties of entanglement.  A modern example is a superposition of motional states of a  mirror \cite{pen-mac}, which in turn spurred a lot of   work on quantum control in nano- and mesoscopic  systems \cite{pen-mac-fol}.

Quantum control schemes involve a lot of   realization-dependent details. This makes it hard to disentangle conceptual issues from the  hardware problems.
In this paper we show how quantum computational circuits \cite{nc,it11} can help in designing and analyzing foundational experiments.

\vspace{-3mm}
\section{Complementarity and control with quantum circuits}
\label{sec:1}
\vspace{-1mm}

Familiar concepts --- ``particle" or ``wave"  ---  represent only one aspect of quantum objects. Although we observe single-photon  interference
(a definite wave-like behaviour), the pattern is produced
click-by-click, in a discrete, particle-like manner \cite{gra86,exp_dc,scully}.
Hence we
adopt, as operational definitions,  the notions of `wave/particle' to stand for
ability/inability to produce interference \cite{it11,green}. As illustrated in Fig.~1(a), these properties are observed using  two mutually exclusive set-ups of the Mach-Zender interferometer (MZI).

 Bohr's complementarity  principle \cite{bohr_wz} ascribes a fundamental significance to this situation. ``The information provided by different experimental procedures that in principle cannot\ldots be performed simultaneously, cannot be represented by any mathematically allowed quantum state of the system" \cite{stapp}.

\begin{figure}[htbp]
\label{delayed_ch}
\includegraphics[width=0.48\textwidth]{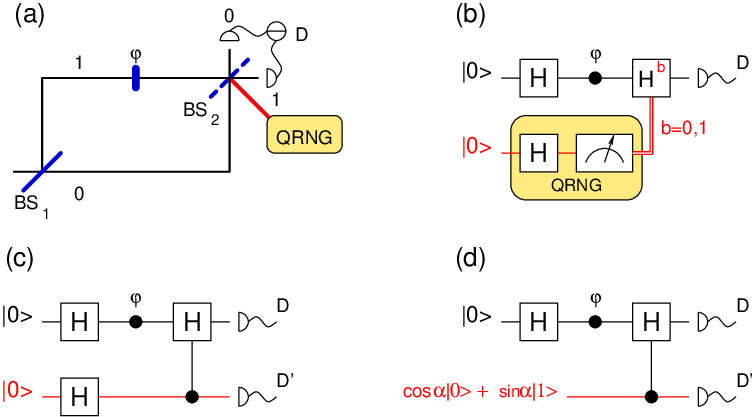}

\caption{\small{ Schematics of the delayed-choice experiments (adapted from \cite{it11})
\newline
(a) Mach-Zender interferometer. A quantum random number generator (QRNG) \cite{exp_dc}
determines weather BS$_2$ is inserted (the output is 1) or not (the output is 0). 
\newline
(b) The equivalent quantum network. An ancilla (red line), initially prepared in the state  $|+\9= (|0\9+ |1\9)/\sqrt{2}$  then measured, acts as QRNG.
\newline
(c) Delayed-choice with a quantum beamsplitter. A quantum device plays the role of the QRNG controling the
Hadamard gate; this makes possible to delay the measurement revealing its output after the application of the
$H$ gate.
\newline
(d) Biasing the QRNG by preparing the ancilla in an arbitrary state $\cos\alpha|0\9+ \sin\alpha|1\9$ is crucial in interpreting the experimental results as supporting wave-particle duality.
}}
\vspace{-5mm}
\end{figure}

Wheeler's delayed-choice   experiment\footnote{It was first discussed by von Weizs\"{a}cker \cite{w-dc} and briefly mentioned by Bohr in his review of the Einstein-Bohr discussions \cite{bohr-dc}.}  \cite{wheeler_dc,leggett}  is designed to eliminate this possibility. As shown in Fig.~1(a) one randomly chooses whether or
not to insert the second beamsplitter only when the photon is already
inside the interferometer and before it reaches BS$_2$.

The rationale behind the delayed-choice is to avoid a possible causal link between the experimental setup and photon's
behaviour: the photon should not ``know" beforehand how to behave.
The choice of inserting
or removing BS$_2$ is  controlled by a random
number generator. 


 A quantum circuit  model \cite{nc} enables us to analyze the {\em gedanken} experiment at a higher level of abstraction and to understand the information flow between different subsystems. The delayed-choice  experiment \cite{it11,lucas,qdc_optics,kaiser12} is equivalent to the quantum network in Fig.~1(b), where Hadamard gates $H$ play the role of beamsplitters; we call the top (black) line {\em the photon} and the bottom (red) line {\em the ancilla}. The quantum random number generator is modelled by an ancilla prepared in the equal-superposition state $\ket{+}= \frac{1}{\sqrt{2}}(\ket{0}+ \ket{1})$, then measured; the result of this measurement (0 or 1) controls if BS$_2$ is inserted or not. Classical control after the measurement of the ancilla in Fig.~1(b) is equivalent to  quantum control before the measurement of the ancilla, Fig.~1(c).

 This seemingly innocuous transformation radically changes the setup and has two profound implications. First, since now we have a quantum beamsplitter in a superposition of being present or absent, the interferometer is in a superposition of being closed or open.
Second, quantum control allows us to reverse the temporal order of the measurements. We can now detect the photon before the ancilla, i.e., before finding out the interferometer is open or closed. This implies that the selection  if the photon behaves as a particle or as a wave is made {\em after} it has been already detected.

Quantum control thus allows us to explore a regime outside the classical realm: in any classically-controlled experiment the choice of inserting or not the second beamsplitter has to be made before the photon is detected. Since the photon and the ancilla interact at the $C(H)$ gate, the ancilla is always prepared before the photon reaches BS$_2$.

In Fig.~1(d), the photon--ancilla system starts in the state $|\Psi\9=\cos\alpha \ket{0}+ \sin\alpha \ket{1}$;   the final state is
\be
\ket{\Psi'}= \cos\alpha \ket{{\sf p }}\ket{0}+ \sin\alpha \ket{{\sf w }}\ket{1},
\label{psi_f}
\ee
where the wavefunctions $\ket{{\sf p}}= \frac{1}{\sqrt{2}}(\ket{0}+ e^{i\varphi}\ket{1})$ and $\ket{{\sf w}}= e^{i\varphi/2}(\cos\frac{\varphi}{2}\ket{0}- i \sin\frac{\varphi}{2}\ket{1})$ describe particle and wave behaviour, respectively. The two states are in general not orthogonal $\braket{{\sf p }}{{\sf w }}= \frac{1}{\sqrt{2}} \cos\varphi$. Eq.~(\ref{psi_f}) implies that if the ancilla is measured to be $\ket{0}$ ($\ket{1}$), the interferometer is open (closed) and the photon behaves like a particle (wave).

\begin{figure}[htbp]
\vspace{-2mm}
\hspace{-2mm}
\includegraphics[scale=0.37]{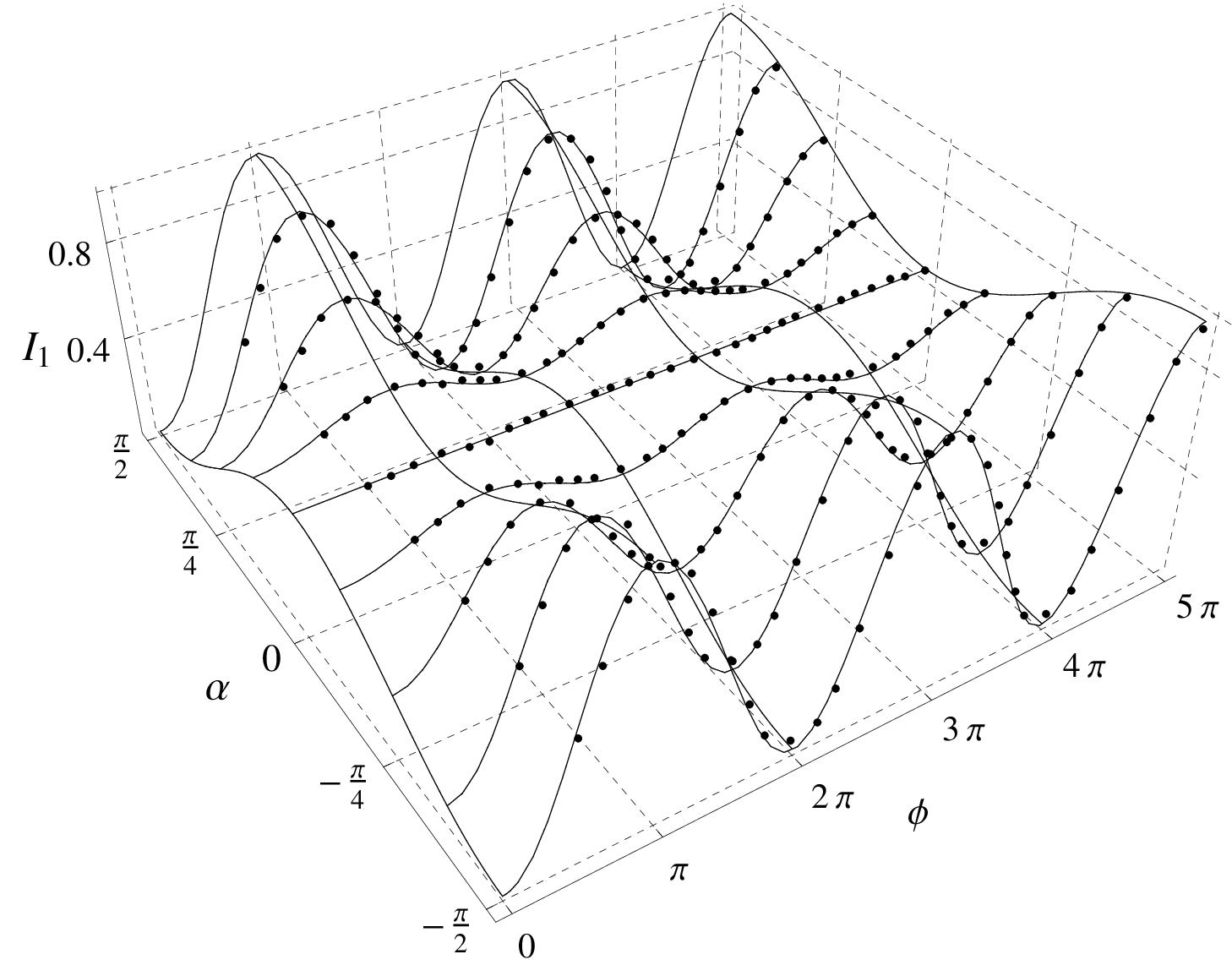} \vspace{-4mm}
 \caption{\small Wave-particle morphing as described by Eq.~(\ref{I0phi_alpha}). The dots represent experimental data adapted from \cite{qdc_optics}. The values $\alpha=\pm\frac{\pi}{2}$ correspond to the wave, and $\alpha=0$ to the particle set-ups of the MZI. The measurements were made in steps of $\pi/8$ in the bias $\alpha$.}
\vspace{-2mm}
\end{figure}

The interference pattern measured by the  {photon detector $D$ is $I_\gamma(\varphi)= \tr(\rho_a \ket{\gamma}\bra{\gamma})$, with $\rho_a= \tr_a\ket{\psi}\bra{\psi}$ the trace over the ancilla in state  $\ket{a}$, and $ \ket{\gamma}=\{ \ket{0}, \ket{1}\}$ denoting the arm of the detector.  If the interferometer is closed, $\ket{a}=\ket{1}$;
 the photon shows  wavelike behaviour with $I_1(\varphi)= I_w(\varphi)= \sin^2\frac \varphi 2$  and visibility $\V= 1$. For an open interferometer $\ket{a}=\ket{0}$; the photon behaves like a particle, with $I_1(\varphi)=I_p(\varphi)=\frac{1}{2}$, resulting in $\V= 0$.
For the ancilla in the superposition given by Fig.~1(b), the reduced density matrix of the photon is $\rho_a= \frac 1 2 (\ket{{\sf p}}\bra{{\sf p}}+ \ket{{\sf w}}\bra{{\sf w}})$, corresponding to $\alpha=\pi/4$ in Eq. (\ref{psi_f}), and
yielding $I_0(\varphi) = \frac{1}{2}+ \frac{1}{4}\cos\varphi$.
  The visibility of the interference pattern is $\V= (I_{max}- I_{min})/(I_{max}+ I_{min})$, where the min/max values are calculated with respect to $\varphi$.}  For the entangled state (\ref{psi_f}) the result is
\be
I_1(\varphi, \alpha)= I_p(\varphi) \cos^2\alpha + I_w(\varphi) \sin^2\alpha.
\label{I0phi_alpha}
\ee
Without correlating the photon data with the ancilla we observe an interference pattern with reduced visibility $\V= \sin^2\alpha$: the photon has a mixed behaviour between a particle and a wave. On the other hand, if we do correlate the photon with the ancilla we observe either a perfect wave-like behaviour (ancilla $\ket{1}$) or a particle-like one (ancilla $\ket{0}$). By varying $\alpha$ we have the ability to modify continuously the interference pattern, morphing from wave to particle patterns (Fig.~2). 

Before discussing the interpretation of this experiment we note that unlike its classically-controlled counterpart, no spacelike separation between the ancilla (taking on the role of a QRNG) and the photon is possible.  We will discuss the consequences of this failure  in Sec.~\ref{sec:anal4}.  

\vspace{-2mm}
\section{Hidden variable models} \label{sec:hv3}
\vspace{-1mm}

How do we know that, for example, the delayed-choice experiment rules out the wave-particle dichotomy? Hidden variables   help to obtain an answer. To this end we  introduce a binary hidden variable $\lambda=\p,\w$ that   represents randomly created photons that are ``really" particles or waves. A more sophisticated construction is discussed in \cite{ijmt12}.

In dealing with  HV theories we  assume the standard conditions for probability distributions; for all variables $i,j$ we have:  $p(i)= \sum_j p(i,j)$  and  $p(i,j)= p(i|j) p(j)= p(j|i) p(i)$. A hidden variable theory should be

(i) adequate, i.e predict the correct quantum probabilities,
\begin{align}\vspace{-1mm}
  & q(a,b,\ldots|A,B,\ldots)   \nonumber \\
  & =\sum_\lambda p(a,b,\ldots|A,B,\ldots,\lambda)p(\lambda|A,B,\ldots),
\end{align}
where $A,B,\ldots$ are measurement set-ups and $a,b,\ldots$ the respective measurement results. For the experiments of Fig.~1  it means
\be \vspace{-1mm}
q(a,b)\equiv p(a,b)= \sum_\lambda p(a| b, \lambda)\, p(b| \lambda)\, p(\lambda). \label{adeq}
\ee
\vspace{-1mm}

Typically  a number of additional assumptions of various strength are made (see \cite{branyan} for their discussion and interrelations). While determinism is one of the key assumptions in the analysis of \cite{ijmt12}, it is not required in dealing with the experiments of Fig.~1. However, we assume that

(ii) a HV model satisfies {$\lambda$-independence} if for all $A, A',B, B',\ldots$
\be
p(\lambda|A,B,\ldots)=p(\lambda|A',B',\ldots),
\ee
where $A$ and $A'$ are two different set-ups of the same measurement. This asserts that the process determining the value of the hidden variable is independent of which measurements are chosen. A spacelike separation in Bell-type or delayed-choice experiments, together with an assumption of absence of superluminal propagation, {are the rationale} for considering this property enforced.


We consider the requirements of ``being a wave" and ``being a particle" as ``real objective properties". This is a specific example of  constraining probability distributions of a HV model to satisfy   particular classical expectations of a system's behaviour.
In this case the HV $\lambda$ determines the behaviour: particle in an open interferometer ($b=0$) and  wave in a closed MZI ($b=1$). Hence

(iii) wave-particle objectivity (or realism) constrains the conditional distributions as
\begin{align}
& p(a|b=0, \lambda=\p)= \left(\half, \half \right),\\
&  p(a|b=1,\lambda=\w)= \left( \cos^2\!\phalf,  \sin^2\!\phalf\right),
\label{p_and_w}
\end{align}
respectively. Note that it is a weaker requirement than determinism, where the knowledge of HV determines the outcomes.

\vspace{-2mm}
\section{Analysis of the delayed-choice experiments} \label{sec:anal4}
\vspace{-1mm}

\textit{Assignment of probabilities.} The behaviour of a wave ($\lambda=\w$) in an open ($b=0$)  and of a particle ($\lambda=\p$) in a closed ($b=1$) interferometer  are unconstrained by (iii). We denote these two unknown distributions by $x$ and $y$, respectively
\begin{align}
& p(a|b=0, \lambda=\w)= (x, 1-x),\\
& p(a|b=1, \lambda=\p)= (y, 1-y).
\end{align}
As we find below, potentially awkward questions about the meaning of $x$ and $y$ do not arise.

We assume that the source randomly and independently emits particle- or wave-like photons with probability $p(\p)=f$ and $p(\w)=1-f$.
The probability $p(a,b,\lambda)$  assignments are completed by using  the conditional probability distributions of the ancilla $b$ and the hidden variable $\lambda$:
\be
p(b| \lambda=\p)=(z, 1-z), \qquad
p(b| \lambda=\w)= (v, 1-v)
\ee
satisfying the consistency condition $p(b)=$  $\sum_\lambda p(b| \lambda) p(\lambda)$.

\textit{Contradiction.} The observed (marginal) probability distribution of the ancilla/QRNG $q(b)$ is
\be
q(b)=p(b)= (\cos^2 \alpha, \sin^2 \alpha).
\ee
Writing explicitly  the adequacy conditions Eq.~(\ref{adeq}) and manipulating the resulting equations we obtain \cite{it11}:
\begin{eqnarray}
v(1-f)(x- \half)&=& 0, \label{3.1} \\
f(1-z)(y- \cos^2 \frac{\varphi}{2})&=& 0, \label{3.2}\\
zf+ v(1-f)- \cos^2\alpha &=& 0. \label{3.3}
\end{eqnarray}

Five of the non-trivial solutions of this system essentially restore wave-particle duality, making behaviour of the photon indpendent of $\lambda$.  The last solution is:
\be
v=0,\ \  z=1,\ \  f=\cos^2 \alpha
\ee
with $x,y$ undetermined. In other words, the source randomly emits particles and waves with a distribution $p(\lambda)= (\cos^2 \alpha, \sin^2 \alpha)$ identical to the probability distribution $q(b)$ of the ancilla, being thus perfectly correlated with MZI being open or closed.

\textit{Implications.} In a classically-controlled delayed-choice experiment   spacelike separation (and the subluminal propagation of signals) enforces $p(b,\lambda)=p(b)p(\lambda)$, i.e., $v\equiv z$ and the last solution is impossible. In this case the conclusion is either a wave-particle duality or deeper conspiratorial correlations (e.g. between $\lambda$ and the settings of QRNG).

In the quantum delayed-choice experiment if  the  ancilla $B$ is considered as part of the measuring device then  a formal conclusion is that \textit{objectivity and $\lambda$-independence are incompatible}.
On the one hand, it can be argued that our result is stronger than the one obtained with a classically-controlled device, since the required correlation is not with the experimental settings (the beam splitter is present or absent), but with the set-up of the random number generator driving it. On the other hand, having a quantum ancilla allows for its hidden variables to somehow possibly compromise the conclusion. Our analysis \cite{ijmt12} demonstrates that this is not the case and provides experimental signatures for the refutation of possible HV theories.

\vspace{-2mm}
\section{Further applications of quantum control}
\vspace{-1mm}

\textit{Entanglement-controlled experiment.} One way to ensure the quantumness of the controlling device is to use the entangled  ancilla \cite{kaiser12}, Fig.~3(b).   Replacing the ancilla by one qubit of a maximally entangled pair, and introducing the bias $\alpha$ into the second half of the pair before it is measured, allows for the ancilla qubit (loosely speaking) to
``not have a state"  before the interaction.

\begin{figure}[htbp]
\vspace{1mm}
\includegraphics[width=0.5\textwidth]{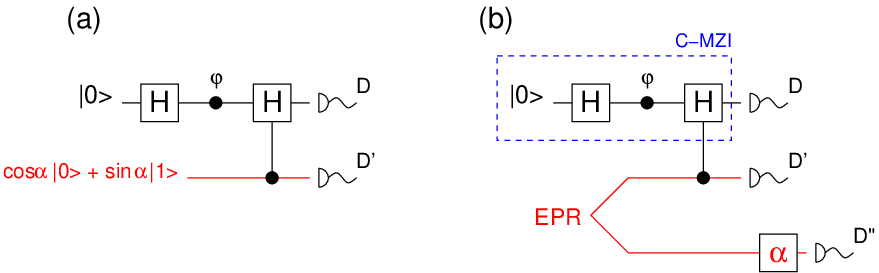} \vspace{-2mm}
\caption{\small Entanglement-controlled QDC
\newline
(a) We bias the quantum random number generator (QRNG) by preparing the ancilla in an arbitrary state $\cos\alpha\ket{0}+ \sin\alpha\ket{1}$.
(b) A pair of maximally entangled qubits replaces a single ancilla. The first qubit serves as a control for the Hadamard gate, while the bias $\alpha$ is introduced to the second.}\vspace{-4mm}
 \end{figure}
The HV description was extended to this set-up in \cite{ijmt12}. It was shown that the intuitive ideas of determinism, wave-particle objectivity and locality are mutually inconsistent.

\textit{Bell-type inequalities.} A quantum control can be used in other experiments as well. For example,
a modification of the CHSH  experiment \cite{chsh,wz,compendum} to include quantum control is rather simple. The gate $X$ creates a desired pair, prepared in a maximally entangled (or perhaps some other) state. On both Alice's and Bob's side the measurement direction ($A$ or $A'A$, $B$ or $B'B$) is chosen not by a random number generator, as in \cite{aspect82}, but by a quantum-controlled gate. Similar to the delayed-choice experiment, the entangled photons are measured before the choices of the directions are made.

On the one hand such a design makes it impossible to talk about the measurement settings determining HV. On the other hand, the entire system including two qubits that Alice and Bob measure and the two ancillas, may be treated as a single entity thus possibly allowing for a consistent HV theory \cite{peres,bell-rmp}.

\begin{figure}[htbp]
\hspace{35mm}
\includegraphics[width=0.3\textwidth]{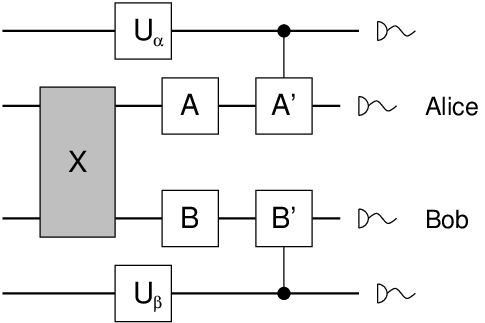}
\caption{\small Quantum-controlled CHSH experiment. The gate $X$ stands for creation of an entangled pair that is shared between Alice and Bob. The controlled $A'$, $B'$ gates are used to select one of the two local measurement set-ups. Type of the measurements Alice and Bob perform is determined only after they detect their ancilae.}\vspace{-1mm}
\end{figure}

\begin{figure}[htbp]
\includegraphics[width=0.53\textwidth]{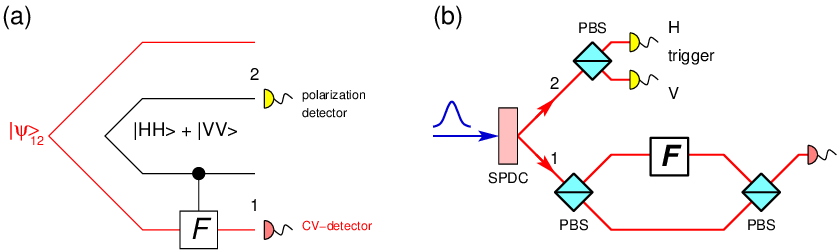}
\caption{\small Optical position and momentum measurement. (a)~Quantum circuit. Two degrees of freedom (transversal coordinate and polarization) of twin photons are used to prepare the hyperentangled state $|\Psi\rangle$. The external lines represent the spatial degrees of freedom while the internal ones stand for the polarization. (b) SPDC can be used to generate   $|\Psi\rangle$. One of the photons is sent through a MZI.  Due to the polarization beamsplitter (PBS) the Fourier transform $F$ is applied conditioned on  polarization.  The actual implementation must use a suitable lens system in order to perform all the necessary operations on the continuous variables  \cite{daniel_PRA}.}
\label{delayed_pm}
\vspace{-4mm}
\end{figure}

\textit{Position and momentum.} Quantum controlled devices can be used in the paradigmatic case of position and momentum observables \cite{bohr_wz}. Consider the specific case of the transversal degrees of freedom of a paraxial and monochromatic light beam \cite{opt,lt}. The study of the electromagnetic field in the quantum regime is based on the quantization of the Fourier components of the classical field, which are analogous to the position and momentum operators of the harmonic oscillator. Regarding the spatial degrees of freedom of the field, a similar quantization of the transversal position ($\mathbf{x}_\perp$) and momentum ($\mathbf{q}_\perp$) variables can be accomplished \cite{longhi,dcc}. The idea here is to get  the information about these complementary observables in a single setup by means of a quantum control, similar to the development of Fig.~\ref{delayed_ch}. This can be done through the circuit shown in Fig.~\ref{delayed_pm}.

Consider a single photon whose transversal spatial profile is a continuous variable quantum system and its polarization plays the role of the ancilla. The experiment starts through the preparation of the ancilla in an equal superposition $\frac{1}{\sqrt 2}(|H\rangle + |V\rangle)$, while the system is prepared in an arbitrary continuous variable state $|\psi\rangle_{S}$ (validity of this approximation is discussed in \cite{lt}). Controlled by the state of the ancilla, a Fourier transform will  or will not be applied to the system, resulting in momentum or position measurements, respectively.  Therefore, focusing on the transversal degrees of freedom only, the final state (before the measurement) can be described as
\begin{equation}
|\Psi'\rangle=N\big(|\psi\rangle_{S}|H\rangle_{A} + |\zeta\rangle_{S}|V\rangle_{A}\big),
\label{estado_5}
\end{equation}
where $\zeta =\mathcal{F}[\psi ]$ is the Fourier transform of $\psi(x)$, $N$ is the normalization constant and $|i\rangle_{A}$ ($i=H,V$) represents the polarization state of the photon. Therefore, using the ancilla qubit to switch between both measurements, we can obtain all the information about the possible measurements of the complementary observables using the same experimental apparatus. Note that the time-ordering of the measurements is not important: we can measure the ancilla before, after or jointly with the photon.

A possible experimental realization of this protocol can be implemented using Spontaneous Parametric Down Conversion (SPDC). In this case, the pump laser is sent through a nonlinear crystal, generating photons with entanglement in the spatial and polarization variables of the field of light. A possible state obtained with this process is the hyper-entangled state (a state with entanglement in more than one degree of freedom):
\begin{equation}
|\Psi\rangle=N|\psi\rangle_{12}\big[|H\rangle_{1}|H\rangle_{2}+|V\rangle_{1}|V\rangle_{2}\big],
\end{equation}
where $|\psi\rangle_{12}$ is the quantum state entangled on the transversal variables of the photons and $|H\rangle_{i}$($|V\rangle_{i}$) is the single photon state of photon $i=1,2$ with horizontal (vertical) polarization. Photon-1 is then sent to the MZI in which the Fourier transform is implemented in one of the arms (see Fig.~\ref{delayed_pm}). Photon-2 is detected and used as a trigger to guarantee that we have a single photon in each run of the experiment. In this way, the complete state before the measurement can be written as
\begin{equation}
|\Psi\rangle=N\big[|\psi\rangle_{12}|H\rangle_{1}|H\rangle_{2} + |\zeta\rangle_{12}|V\rangle_{1}|V\rangle_{2}\big].
\label{final}
\end{equation}
$|\zeta\rangle_{12}$ represents the state of both photons after the implementation of the Fourier transform on photon-1. If we measure photon-2 in a specific position and scan photon-1 detector over the transversal direction, the result of the coincidence counting will be the probability distribution of photon-1 in transversal position or transversal momentum, depending of the polarization of photon-2. Therefore, the correlations between these measurements contain all the information about both complementary variables, as can be seen directly from Eq. (\ref{final}).

\vspace{-2mm}
\section{Summary} \label{sec:exper}\vspace{-1mm}

The use of quantum control necessitates a reassessment of Bohr complementarity \cite{ns}. Partial information about complementary quantities can be obtained in a single experiments \cite{scully}. Contrary to Bohr's opinion, we do not have to change the experimental setup in order to measure complementary properties \cite{it11} --- we can measure both properties in a single experiment, provided that a component of the apparatus is a quantum object in a superposition state. The behaviour is post-selected by the experimenter after the photon has been detected, by correlating the data with the appropriate value of the ancilla.

A quantum control makes impossible the spacelike separation between the device settings and the system. A spacelike separation can be reintroduced by having  an additional classical device or creating sub-systems at  mutually spacelike events \cite{ijmt12}. Quantum delayed-choice experiments can force  proponents of   wave-particle objectivity to accept the same level of conspiracy as their classically-controlled counterpart experiments do with  spacelike separation. Adding  spacelike separation between the photon and the ancilla leads to new results \cite{ijmt12}. However, it is still not clear to what extent its introduction is necessary in other experiments.


One of the consequences of the quantum control is the morphing between wave and particle statistics, see Eq.~(\ref{I0phi_alpha}). The measurements reported in ref.~\cite{qdc_optics} are in excellent {agreement} with the theoretic prediction. 

Quantum control showed its usefulness in the delayed-choice experiments. As it should be clear from our presentation, it can be used in any experiment where several alternative set-ups are employed. We expect to see both  conceptual surprises and practical benefits stemming from its use.

\begin{center}
\textbf{ACKNOWLEDGMENTS}
\end{center}\vspace{-2mm}

R.G. and L.C. thank CAPES, CNPQ and INCT-IQ for partial financial support.
T.J. and R.B.M. were supported in part by the NSERC of Canada.
D.R.T. thanks the Center for Quantum Technologies  at the National University of Singapore for hospitality, Berge Englert,  Valerio Scarani and Vlatko Vedral for useful discussions and Alla Terno for the help with visualizations. We are grateful to Chuan-Feng Li and Jian-Shun Tang for kindly sharing their data with us.

\end{document}